\documentclass[conference]{IEEEtran}
\IEEEoverridecommandlockouts
\usepackage{cite}
\usepackage{amsmath,amssymb,amsfonts}
\usepackage{algorithmic}
\usepackage{graphicx}
\usepackage{textcomp}
\usepackage{siunitx}
\usepackage{xcolor}
\usepackage{float}
\usepackage{dblfloatfix}
\def\BibTeX{{\rm B\kern-.05em{\sc i\kern-.025em b}\kern-.08em
    T\kern-.1667em\lower.7ex\hbox{E}\kern-.125emX}}

\begin{document}

\title{Benchmarking and modeling of analog and digital SRAM in-memory computing 
architectures

}

\author{\IEEEauthorblockN{Pouya Houshmand}
\IEEEauthorblockA{\textit{MICAS, KU Leuven} \\
Leuven, Belgium \\
pouya.houshmand@kuleuven.be}
\and
\IEEEauthorblockN{Jiacong Sun}
\IEEEauthorblockA{\textit{MICAS, KU Leuven}\\
Leuven, Belgium \\
jiacong.sun@kuleuven.be}
\and
\IEEEauthorblockN{Marian Verhelst}
\IEEEauthorblockA{\textit{MICAS, KU Leuven} \\
Leuven, Belgium\\
marian.verhelst@kuleuven.be}
}

\maketitle

\begin{abstract}
In-memory-computing is emerging as an efficient hardware paradigm for deep neural network accelerators at the edge, enabling to break the memory wall and exploit massive computational parallelism. Two design models have surged: analog in-memory-computing (AIMC) and digital in-memory-computing (DIMC), offering a different design space in terms of accuracy, efficiency and dataflow flexibility. This paper targets the fair comparison and benchmarking of both approaches to guide future designs, through a.) an overview of published architectures; b.) an analytical cost model for energy and throughput; c.) scheduling of workloads on a variety of modeled IMC architectures for end-to-end network efficiency analysis, offering valuable workload-hardware co-design insights.
\end{abstract}


\section{Introduction}
The need for executing AI workloads on edge devices, characterized by limited power budgets and area constraints, results in large research interest in custom hardware accelerator design, with large advancements in a short time frame \cite{edge_intelligence}.
As these workloads require a significant amount of data movement between the processor and the memory, traditional Von Neumann architectures have proven unsuitable for the application \cite{intro_2}.
To handle this data transfer bottleneck, a functional approach adopted from the early published designs is to parallelize the computing units in space, thus allowing for a higher degree of spatial multi-cast reuse of the operands as they are fetched from memory. 
Conventional accelerators \cite{intro_3} thus are composed of a 2D array of processing elements (PE), each of which contains a multiply-accumulate (MAC) unit (or vector) and small register files to store operands.
These operands are fetched from larger memories outside the PE array and distributed in space across the processing elements, according to the spatial parallelization scheme adopted for the workload (also called "spatial unrolling") \cite{zigzag}. Yet, memory interfacing remains a performance-limiting aspect of modern neural accelerators.
In recent years, in-memory computing (IMC) has therefore emerged as a promising alternative to PE-based accelerators, by performing the MAC operations near/in the memory cells directly. This allows to greatly reduce access overheads and enables massive parallelization opportunities, with potential orders of magnitude improvements in energy efficiency and throughput \cite{murmann_aimc}.
Most recent IMC designs published in the literature are focused on analog IMC (AIMC), where the computation is carried out in the analog domain. While this approach ensures extreme energy efficiencies and massive parallelization, the analog nature of the computation and the presence of intrinsic circuit noise and mismatches compromises the output accuracy. Furthermore, the rigid structure of the computation's dataflow limits the spatial mapping possibilities. 
To avoid the hurdles of AIMC, digital in memory computing (DIMC) is lately gaining more interest as a valid alternative, thanks to its noise-free computation and more flexible spatial mapping possibilities, trading off added flexibility and accurate computation for less energy efficiency.
These new opportunities stemming from AIMC and DIMC resulted in many recent implementations and publications in the literature. However, these works vary strongly in terms of hardware architecture, array dimensions, and silicon technology. This makes it difficult to grasp their relative strengths, trends and future directions.
While several works assess and discuss IMC trends qualitatively
\cite{imc_trends_1, imc_trends_2, burr_benchmarking, murmann_aimc, imc_trends_3, imc_trends_4, imc_trends_6, imc_trends_7}, only few aim at quantitatively modeling or benchmarking architectural strategies.  
The latter, moreover, primarily focus on AIMC designs \cite{imc_modeling_1, imc_modeling_2, murmann_aimc, imc_modeling_3, imc_modeling_4, imc_modeling_5, imc_modeling_6, imc_modeling_8}, while there is a lack of DIMC modeling efforts. Similarly, for mapping space explorations, most of the focus has been dedicated to AIMC designs, while lacking DIMC assessment \cite{imc_dse_1, imc_modeling_4, imc_modeling_5, imc_dse_2}.

To this purpose, this paper aims at:
\begin{itemize}
    \item Providing a comprehensive assessment of recently published AIMC and DIMC chips, in order to understand through benchmarking 
    the capabilities and limitations of the two design paradigms.
    \item Provide a unified analytical model for IMC architectures, validated against numerous design points from literature. 
    \item Integrate the model into ZigZag \cite{zigzag}, a design space exploration (DSE) framework for hardware accelerators. This allows to compare different AIMC / DIMC hardware design points, data flows and mappings in terms of energy and throughput efficiency, towards insights on optimal design points for targeted tinyMLperf workloads \cite{mlperf}.
\end{itemize}


\begin{figure}[!b]
    \centering
    \includegraphics[ width=\linewidth]{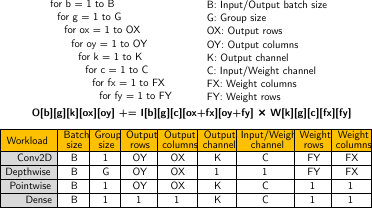}\\
    \vspace{0.1in}
    \includegraphics[width=\linewidth]{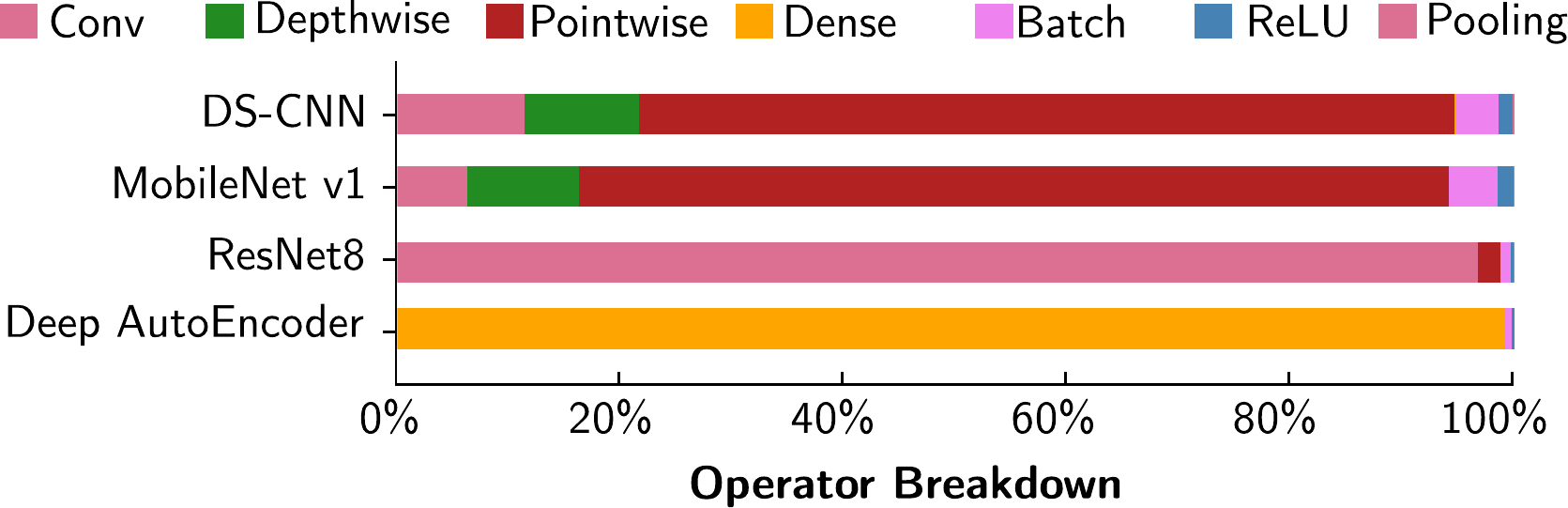}
    \caption{8-nested loop DNN layer representation, workloads representation and operator breakdown of tinyMLPerf \cite{mlperf} benchmark models}
    \label{fig:fig_1}
\end{figure}

\section{Background}
\label{sec:02_background}

\subsection{Dataflow concepts for IMC}
\begin{figure}[!t]
    \centering
    \includegraphics[width=\linewidth]{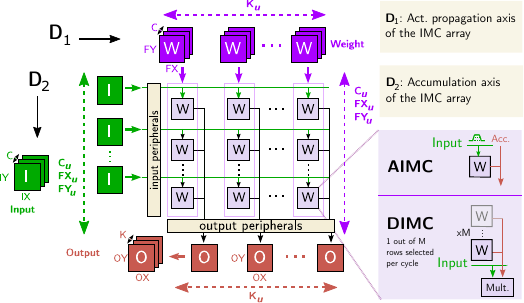}
    \caption{In-memory computation paradigm}
    \label{fig:fig_3}
\end{figure}

Deep neural network (DNN) workloads consist of a sequence of layers. The operations in the most common layers can be described as a combination of 8-nested for loops, which iterate over the indices of an input feature map tensor $I$, a weight tensor $W$ and that generate an output tensor $O$, as shown in Fig. \ref{fig:fig_1}.  
The nested loop representation opens opportunities for parallelization of the operations, by means of \textit{tiling} and \textit{spatially unrolling} the loops across an array of computational units \cite{murmann_aimc}.
The adopted spatial mapping heavily impacts the system-level performance of the accelerator, as it contributes considerably to the access patterns of the operands. Multi-casting the operands across one or more dimensions of the computational array reduces the memory accesses to higher levels in the memory hierarchy by a factor proportional to their spatial reuse, which translates in energy and latency savings at system level.
To achieve this, the tensor operations can be decomposed in a sequence of matrix-vector multiplications (MVM) by tiling the suitable loops \cite{quant_survey}.
MVM operations offer a great opportunity for in-memory acceleration as their dense 2D array structure aligns well with the array structure of the memory macros. 
Due to the physical properties of the hardware template, the spatial unrolling dimensions of the weight matrix are chosen so as to maximize spatial reuse of the activations along the columns of the memory, and accumulation of the partial sums along the rows of the IMC array, as shown in Fig. \ref{fig:fig_3}: The $K$ loops -- irrelevant for the inputs -- are typically unrolled across the columns, while the $C$, $FX$, $FY$ loops -- irrelevant for the outputs -- are parallelized across the rows (Fig \ref{fig:fig_3}). To further speed up computation and increase the reuse of operands at chip-level, the $OX$, $OY$ or $G$ loops can be parallelized across multiple IMC macros on the same die, requiring, however, duplication of the weights \cite{aimc_2}. 

\subsection{SRAM-based IMC}
SRAM is largely deployed for IMC, thanks to its robustness, large scale integration in CMOS, high endurance, and reliability, when compared to NVM-based solutions. This makes it largely preferred for small to medium capacity systems, even though it does not guarantee as high densities as its NVM counterparts.

IMC designs can be categorized into AIMC and DIMC. In DIMC macros, the weight values that are stored in one selected row are read out and directly combined with an input vector for digital multiplication, in a binary multiplier cell which is tightly integrated with the memory cells. Multiplication results are further accumulated in the digital domain with an adder tree. In AIMC, on the other hand, all rows are jointly activated, with each SRAM cell performing one multiplication of a binary weight and a given input along each word line. AIMC guarantees higher computational density and better energy efficiencies, but this only if 1.) the peripheral cost of A/D and D/A conversions are amortized across a very large array and 2.) a high utilization of the IMC array is achieved \cite{iedm_aimc}. 
While DIMC has less intrinsic efficiency, 
it typically comes with more flexibility in the IMC array structure: Since it is not burdened by A/D peripherals, it can easily be reconfigured into more granular and smaller arrays.
To compare the two design paradigms, a benchmarking survey is presented in the next section.


\section{Benchmarking}
\label{sec:03_benchmarking}


While it would be ideal to benchmark all AIMC and DIMC designs based on their capabilities on actual workloads with metrics such as energy/inference and accuracy, most publications report only peak energy efficiency and throughput figures at the IMC macro level to highlight the capabilities of their designs. Therefore, in this paper, we will firstly focus on bank-level benchmarking of IMCs, after which modeling will expand towards full workload efficiency implications.

\begin{figure*}[!b]
    \centering
    \includegraphics[width=\linewidth]{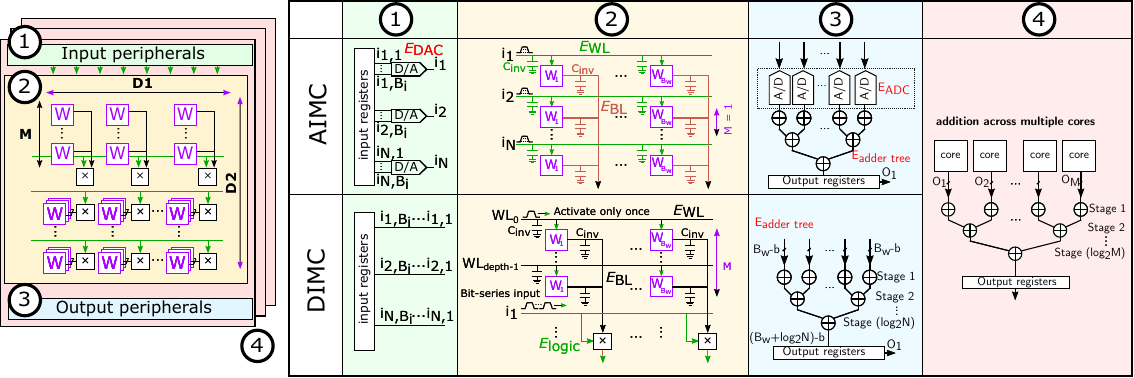}
    \caption{Modeling template for AIMC and DIMC architectures}
    \label{fig:imc_model}
\end{figure*}

\begin{figure}[t]
    \centering
    \includegraphics[width=\linewidth]{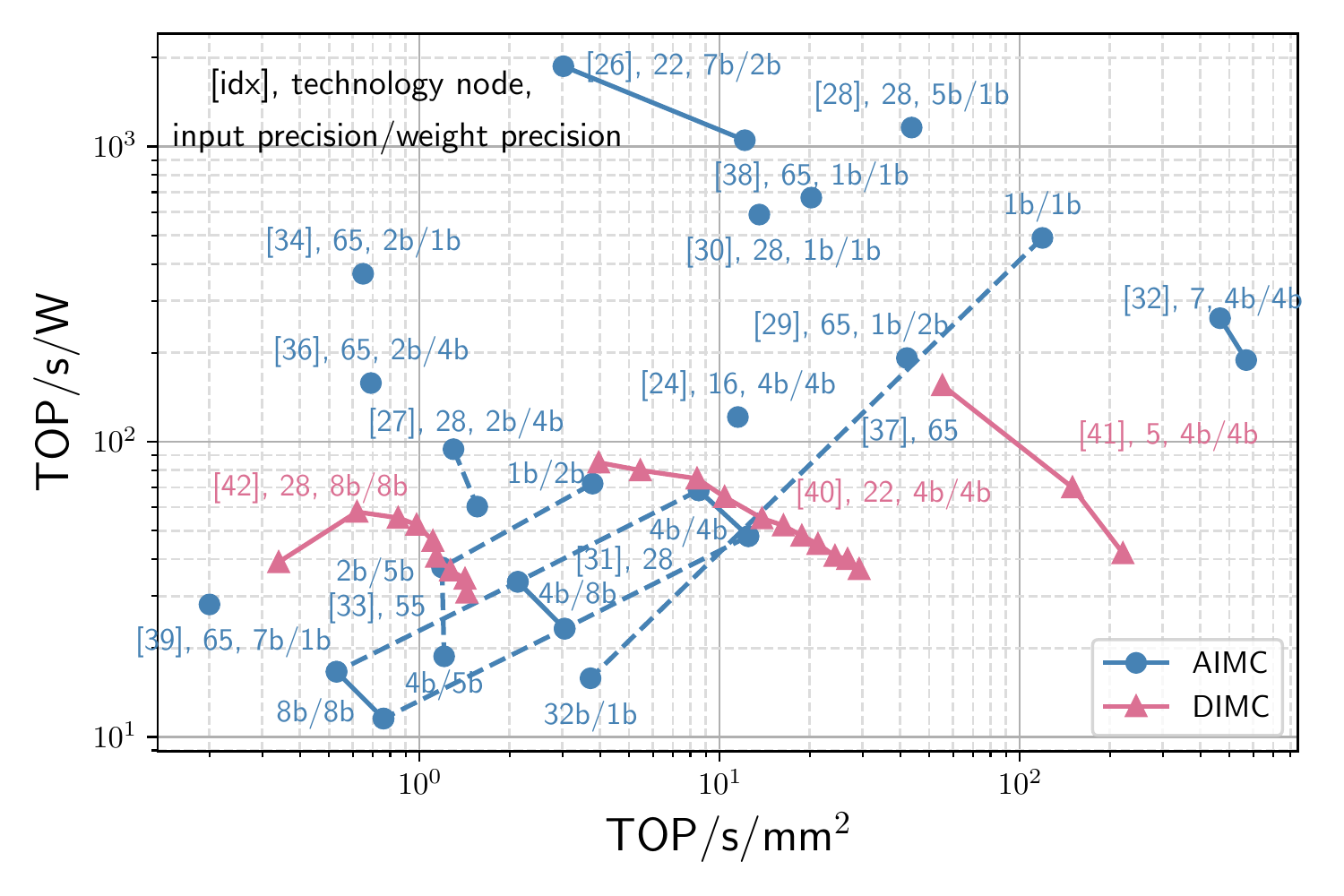}
    \caption{Benchmarking on AIMC/DIMC architectures. Each point reports the technology node deployed for the design and the input/weight bit precision used.}
    \label{fig:benchmarking_fig_1}
\end{figure}

To quantitatively compare the IMC designs reported in literature, we collect those works that target MVM operations and non binary neural networks (BNNs). The BNN restriction is due to the fact that 1) for AIMC the A/D conversion is substituted with an energy efficient sense-amplifier (at the cost of added quantization noise), skewing the efficiencies comparison and 2) the tinyMLPerf benchmark \cite{mlperf} does not include BNN models. Only IMC works 
that report their performance with 50\% sparsity 
are retained to guarantee fairness in the comparison. This results in following selected AIMC designs: \cite{aimc_1,aimc_2,aimc_3,aimc_4,aimc_5,aimc_6,aimc_7,aimc_8,aimc_9,aimc_10,aimc_11,aimc_12,aimc_13,aimc_14,aimc_15}; and DIMC designs: \cite{dimc_1,dimc_2,dimc_3}. 
 For each implementation, Figure \ref{fig:benchmarking_fig_1} plots the derived energy efficiency (TOP/s/W) and computational density (TOP/s/mm$^2$) from the papers' reported peak efficiencies. Dashed, resp. solid lines connect different results stemming from the same chip for across different operand precision, resp. supply voltages. 
Other works (\cite{imc_trends_1, survey_sscm}) benchmark different designs on bit-normalized metrics (e.g. 1-b TOPs/W, 1-b TOP/s); this work does not employ such metrics as they would provide a distorted representation of the design landscape. 
The impact of operand precision will become more clear in the modeling Sec. \ref{sec:04_imc_modeling}. 

Fig. \ref{fig:benchmarking_fig_1}, reveals that \cite{aimc_1} achieves the best peak energy efficiency of $\sim$1800 TOP/s/W among AIMC designs. This is possible through the optimized design of ADCs and DACs and a large array that amortizes their costs -- making the array power consumption the dominant contribution instead of the ADCs. The best computational density is achieved by \cite{aimc_8}, a design implemented in 7nm with Flash ADCs, as opposed to the commonly used SAR-ADCs; this design choice in turn affects the energy efficiencies, which are not optimal. Multi-core architectures have also emerged recently to reduce the lack of flexibility in AIMC dataflows by either unrolling the OX dimension across macros \cite{aimc_5} or using smaller arrays \cite{aimc_6, aimc_7, aimc_8, aimc_10}, with benefits both in terms of utilization and signal margin on the ADCs. It is to be noted that in AIMC designs, the technology node plays a role in achieving a high area density, but does only marginally affect energy efficiency. On the other hand, the performance of DIMC is highly dependent on the technology node and operand precision: smaller technology nodes guarantee higher computational densities with equal precision and improved power consumption (as in \cite{dimc_2} and \cite{dimc_3}), while higher precisions cause drops in computational density with similar technology (as in \cite{dimc_1} and \cite{dimc_2}).

The interplay between technology node, precision, hardware architecture and operating point determines the final efficiencies. It is however hard to discern these relationships based on peak TOPs/W, TOP/s/mm$^2$ values alone. To assess the impact of each design parameter, a more in depth modeling effort is required. For this, a unified cost model for DIMC and AIMC is presented in the following section.

\section{Analytical modeling for IMC}
\label{sec:04_imc_modeling}

\begin{table}[]
\begin{tabular}{ll}
\multicolumn{2}{c}{\textbf{Hardware model parameters}}                                                                                             \\ \hline
\multicolumn{1}{|l|}{\textbf{R, C}}                                   & \multicolumn{1}{l|}{IMC array row, columns resp.}                            \\ \hline
\multicolumn{1}{|l|}{\textbf{ADC$_{\text{res}}$, DAC$_{\text{res}}$}} & \multicolumn{1}{l|}{Bit resolution of the ADC, DAC resp.}                    \\ \hline
\multicolumn{1}{|l|}{\textbf{WL, BL}}                                 & \multicolumn{1}{l|}{SRAM Wordline, Bitline resp.}                            \\ \hline
\multicolumn{1}{|l|}{\textbf{G$_{\text{MUL}}$, G$_{\text{FA}}$}}      & \multicolumn{1}{l|}{Number of gates per 1-b multiplier, full adder resp.}    \\ \hline
\multicolumn{1}{|l|}{\textbf{M}}                       & \multicolumn{1}{l|}{Number of memory rows multiplexed per vector MAC} \\ \hline
\multicolumn{1}{|l|}{\textbf{B$_w$}}                                  & \multicolumn{1}{l|}{Weight bits stored in parallel in the IMC}               \\ \hline
&\\

\multicolumn{2}{c}{\textbf{Derived parameters}}                                                                                                      \\ \hline
\multicolumn{1}{|l|}{\textbf{D$_1$}}                           & \multicolumn{1}{l|}{Activation propagation axis size of the IMC array}               \\ \hline
\multicolumn{1}{|l|}{\textbf{D$_2$}}                           & \multicolumn{1}{l|}{Accumulation axis size of the IMC array}               \\ \hline
\multicolumn{1}{|l|}{\textbf{N, B}}                                   & \multicolumn{1}{l|}{Adder tree number of inputs, input precision resp.}      \\ \hline
\multicolumn{1}{|l|}{\textbf{F}}                           & \multicolumn{1}{l|}{Total number of 1-b full-adders}               \\ \hline
\multicolumn{1}{|l|}{\textbf{C$_{\text{gate}}$}}  & \multicolumn{1}{l|}{Capacitance per gate}                                    \\ \hline
&\\
\multicolumn{2}{c}{\textbf{Mapping dependent extracted parameters}}                                                                                                    \\ \hline
\multicolumn{1}{|l|}{\textbf{CC$_{\text{prech}}$}}                        & \multicolumn{1}{l|}{Precharging cycles on the bitlines}                                                        \\ \hline
\multicolumn{1}{|l|}{\textbf{CC$_{\text{acc}}$}}                    & \multicolumn{1}{l|}{Cycles of digital accumulation across the adder tree}                                                        \\ \hline
\multicolumn{1}{|l|}{\textbf{CC$_{\text{BS}}$}}                            & \multicolumn{1}{l|}{Number of complete DAC conversions required}                                                        \\ \hline
&\\
\multicolumn{2}{c}{\textbf{Technology dependent fitted model parameters}}        \\                                                                              \hline
\multicolumn{1}{|l|}{\textbf{C$_{\text{inv}}$}}    & \multicolumn{1}{l|}{Capacitance of an inverter, gate resp.}                  \\ \hline
\multicolumn{1}{|l|}{\textbf{k$_1$}, k$_2$}                           & \multicolumn{1}{l|}{ADC model constants from \cite{murmann_aimc}}                                      \\ \hline
\multicolumn{1}{|l|}{\textbf{k$_3$}}                                  & \multicolumn{1}{l|}{DAC energy per conversion step constant}                          \\ \hline

\end{tabular}
\caption{Acronym table}
\label{tab:acronym}
\end{table}

\subsection{A unified energy model for IMC}
To enable a fair high-level comparison between AIMC and DIMC, a unified cost model should include the energy contributions of the most dominant components for AIMC and DIMC. Our model is illustrated in Fig. \ref{fig:imc_model}. It aims at modeling the datapath components of A/DIMC, while reading and writing from higher-level memories for inputs and outputs access will later be accounted for through integration of the model into the ZigZag DSE framework. The total datapath energy $E_{\text{total}}$ can hence be computed as:
\begin{equation}
E_{\text{total}} = E_{\text{MUL}} + E_{\text{ACC}} + E_{\text{peripherals}}  
\end{equation}
where the total energy $E_{\text{total}}$ is computed as the sum of the multiplication energy done in the IMC array ($E_{\text{MUL}}$), the energy cost of digital accumulation outside the IMC array ($E_{\text{ACC}}$) and the contribution of peripherals required to access the array and provide the data ($E_{\text{peripherals}}$). All acronyms for symbols in our model are listed in Table \ref{tab:acronym}.



\subsection{Multiplication cost}
In IMC, the multiplications of the MVM operation occurs within the memory array. The values of the input vector are propagated along wordlines and  combined with the data stored in the memory cells (Fig. \ref{fig:imc_model}.2). The energy required to execute this operation $E_{\text{MUL}}$ can be expressed in the unified model as a sum of two components (Fig. \ref{fig:imc_model}.2): the energy related to the charge/discharge of internal capacitances related to the memory cell operations ($E_{\text{cell}}$) and the energy associated with the extra digital logic required to carry out the multiplication ($E_{\text{logic}}$). The latter is only accounted for in DIMC designs, as in AIMC, the multiplication operation takes place in the operation of the memory cell itself. 

\begin{equation}
    E_{\text{MUL}} = E_{\text{cell}} + E_{\text{logic}}
\end{equation}
 \subsubsection{$E_{\text{cell}}$ contribution}
 $E_{cell}$ stems from charging and discharging the capacitances across the SRAM wordline (WL) and bitline (BL) associated with a read operation. 
For AIMC, the values on the bitlines change every computation cycle since a different input vector is combined with the weights in the array each time. This is not always the case for DIMC: with bit-parallel/bit-serial (BPBS) computations in DIMC (\cite{dimc_1,dimc_2, dimc_3}) the weight values are kept stationary, while different input bits are serially combined across separate cycles. 
To model this, Eq. \ref{eq:eq_3} scales wth the charging cycles ${\text{CC}_{\text{prech}}}$, being the number of cycles in which the values on the bitlines of the IMC array are non-stationary. 


\begin{equation}
    \label{eq:eq_3}
    E_{\text{cell}} = (E_{\text{WL}} + E_{\text{BL}})  \times \text{CC}_{\text{prech}}
\end{equation}
The WL energy $E_{\text{WL}}$ is further computed as in Eq. \ref{eq:eq_4}, proportional to the wordline capacitance per cell $C_{\text{WL}}$ -- assumed to be similar to the capacitance of an inverter in the same technology ($\sim C_{\text{inv}}$) --, the number of weight bits per operand $B_w$ and $D_1$, which is the number of operands per memory row. 

\begin{equation}
    \label{eq:eq_4}
    E_{\text{WL}} = C_{\text{WL}} V^2 B_{w} \times D_1
\end{equation}

Similarly, the BL energy $E_{\text{BL}}$ is estimated as in Eq. \ref{eq:eq_5}, 
with $D_2$ being the accumulation dimension of the IMC array and M the row multiplexing factor. 


The parameter M identifies the number of rows multiplexed per vector multiplication: in AIMC designs this values is equal to 1, since each row does a vector multiplication in each cycle; for DIMC and near-memory compute (NMC) designs not all rows are activated at once per multiplication, as in \cite{dimc_2} 

\begin{equation}
    \label{eq:eq_5}
    E_{\text{BL}} = C_{\text{BL}} V^2 B_{w} \times D_2 \times \text{M}
\end{equation}

 \subsubsection{$E_{\text{logic}}$ contribution}
 

In DIMC designs, the bit-parallel/bit-serial (BPBS) operation is performed, where input bits are serially multiplied with parallel weight bits in the array. The multiplier is typically a single NAND/NOR gate per cell.
We therefore use a unified model for the logic cost:

\begin{equation}
    \label{eq:eq_8}
    E_{\text{logic}} =  V^2 C_{\text{gate}} G_{\text{MUL}} \times \text{total MACs}
\end{equation}

with $C_{\text{gate}}$ being the capacitance for standard logic gate ($\approx 2\times C_{\text{inv}}$). $G_{\text{MUL}}$ is the total number of gates per multiplier; in the case of a bit-serial input it equals to the numbers of gates of a 1-b multiplier ($\simeq 1$) times the weight precision $B_{w}$. 

\begin{figure*}[!b]
    \centering
    \includegraphics[width=\linewidth]{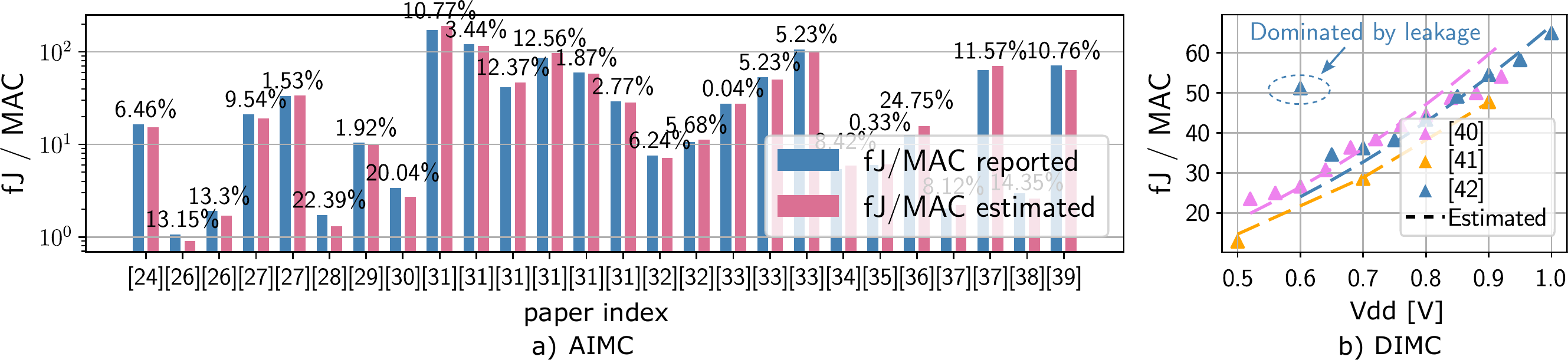}
    \caption{IMC model validation}
    \label{fig:imc_validation}
\end{figure*}

\subsection{Accumulation cost}



Accumulation in the analog domain occurs as a results of charge sharing/injection on each bitline across the memory cells connected to it, which in turn results in a charge and voltage variation over the line (Fig. \ref{fig:fig_3}). This contribution is already captured in Eq. \ref{eq:eq_5}. The analog value then has to be converted by ADC in the digital domain (E${_\text{ADC}}$) and eventually combined with other partial results with a digital adder (E${_\text{adder tree}}$). 
Digital IMC avoids the need for an ADC, substituted by accumulation in the digital domain (Fig. \ref{fig:imc_model}.3). These contributions are accounted for in Eq. \ref{eq:eq_9}.

\begin{equation}
E_{\text{ACC}} = E_{\text{ADC}} + E_{\text{adder tree}}    
\label{eq:eq_9}
\end{equation}

\subsubsection{ADC energy} E${_\text{ADC}}$ 
is estimated using the ADC energy model from \cite{murmann_aimc} 
\begin{equation}
    E_{\text{ADC}} = (k_1 \times \text{ADC}_{\text{res}} + k_2 4^{\text{ADC}_{\text{res}}}) V^2 \times B_w \times (\text{total MACs} / D_2)
\end{equation}

with $k_1 = \SI{100}{\femto\joule}$ and $k_2 = \SI{1}{\atto\joule}$. The model assumes one ADC conversion per bitline; this is true for most designs, except for \cite{aimc_8}, where a 4-bit Flash ADC is used every 4 BLs.

\subsubsection{Adder tree energy}
The digital accumulation energy $E_{\text{adder tree}}$ is modeled as:

\begin{equation}
    \label{eq:eq_6}
    E_{\text{adder tree}} = C_{\text{gate}}G_{\text{FA}}V^2 \times D_1 \times \text{F} \times \text{CC}_{\text{acc}}
\end{equation}

where $C_{\text{gate}}$ the gate capacitance for a standard logic gate, $G_{\text{FA}}$ the number of gates of a 1-b full adder (FA), assumed to be 5. Assuming a ripple carry-adder, the total number of 1-b full adders $F$ executed per cycle per output channel can be derived as:
\begin{equation}
\begin{split}
 \text{F} &= \sum^{\log_2N}_{n} (B + n - 1) \frac{N}{2^n}\\ 
 &= B N + N - B + \log_2 N - 1
\end{split}
\end{equation}

where N is the number of inputs of the first stage of the tree and B the input precision of the adder operands. Since digital accumulation occurs across core rows for DIMC, $N$ is equal to $D_2$ for DIMC, while $N = B_{w}$ for AIMC since accumulation occurs across adjacent bitlines if weight bits are parallelized. $B$ is equal to $B_{w}$ for DIMC, while for AIMC it is ADC$_{\text{res}}$.

\subsection{Peripherals cost}
While for DIMC designs, the drivers propagating the activation values across the wordlines have negligible overhead besides the one already accounted for WL capacitance, multi-bit AIMC accelerators necessitate D/A converters to carry out the task. To this purpose, 
$E_{\text{peripherals}}$ covers the DAC conversion energy contribution and can be expressed as E$_{\text{DAC}}$ (Fig. \ref{fig:imc_model}.1)
\begin{equation}
    E_{\text{DAC}} = k_3 \text{DAC}_{\text{res}} V^2 \times \text{CC}_{\text{BS}} 
\end{equation}
with $k_3 \sim \SI{44}{\femto\joule}$ and CC$_{BS}$ total number of cycles required for activation bits to be converted to the analog domain.

\begin{figure}[t]
    \centering
    \includegraphics[width=\linewidth]{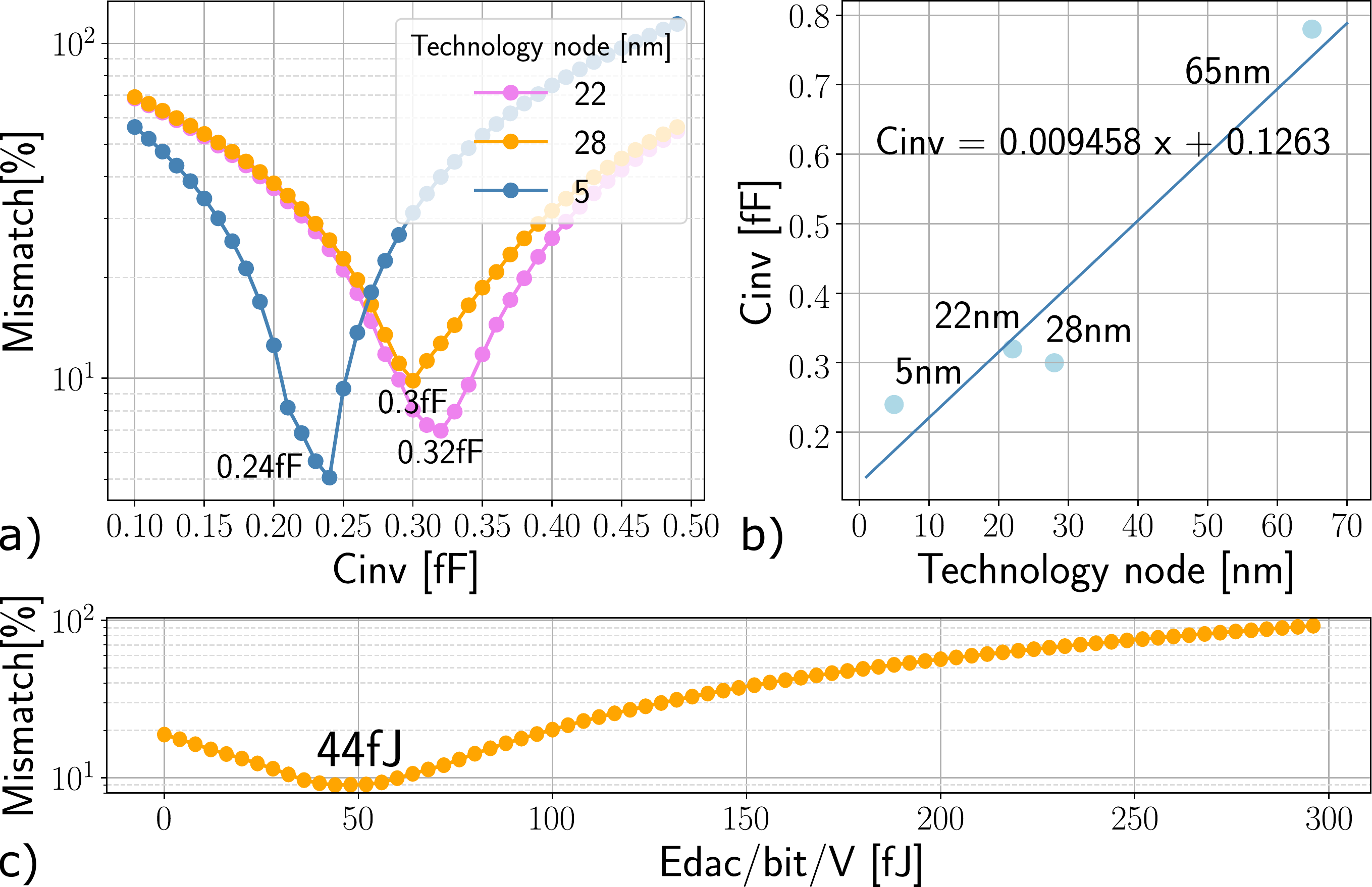}
    \caption{Technology dependent parameters extraction}
    \label{fig:interpolation}
\end{figure}

\subsection{Technology dependent parameters extraction}

The implementation technology impacts the different gate, WL, BL capacitance vales of the unified model. To further unify the modeling, all capacitance values are related to a reference capacitance $C_{inv}$. 
The $C_{inv}$ values for our model across technologies is linearly regressed across fitted $C_{inv}$ values of the different published
DIMC designs (\cite{dimc_1, dimc_2, dimc_3} and \cite{problp} (Fig. \ref{fig:interpolation}.a and \ref{fig:interpolation}.b).
A mismatch between the model and the reported data of $\sim$10\% is observed, stemming from 1) modules not taken into account in the model and 2) leakage power, especially under a low voltage and frequency. 




The same interpolation principle applies to DAC modeling, to derive a 
DAC fJ/conversion ($E_\text{DAC/bit}$) parameter. As shown in Fig. \ref{fig:interpolation}.c, by setting $E_\text{dac/bit}$ to $k_3=\SI{44}{\femto\joule}$, we achieve an average modeling mismatch of $\sim$9\% across all AIMC data points used for modeling.

\section{Validation}

The model has been validated against all design points presented in Sec. \ref{sec:03_benchmarking}. For AIMC (Fig. \ref{fig:imc_validation}.a) mismatches between the model and the reported values are within 15\% for most designs. Large mismatches are often due to unaccounted overheads or inefficient designs: in \cite{aimc_4, aimc_5, aimc_12} reported ADC energies are larger ($\sim4\times$ than the ones estimated by the model). Other notable mismatches arise when large digital overheads are present in the macros, such as in \cite{aimc_6,aimc_12}. For DIMC (Fig. \ref{fig:imc_validation}.b) the model matches closely with the reported values, assuming 50\% sparsity on the operands; the analytical model however does not cover the leakage contribution, which becomes dominant at low voltages and low frequencies: this can be observed for \cite{dimc_3} where measured values at 0.6V steeply diverge from the estimations.

\section{Case studies}
\label{sec:05_experiments}

Peak performances reported in literature and Sec. \ref{sec:03_benchmarking} are not representative for comparing different designs. Instead, architectures should be compared on real workloads to prove their effectiveness. Targeting edge applications, we will compare selected IMC designs on the tinyMLPerf models \cite{mlperf} to assess how AIMC and DIMC 
array sizes affect mapping efficiency. For this first study, we select from the modeled and validated designs, four 
designs in the same technology node and with the same operand precision. A summary of the features of the selected designs is reported in Table \ref{tab:exp_1}. While the macro array size is the same as the one reported on the papers, the number of macros is scaled to make all designs have the same total number of SRAM cells (the size of the largest design) for a fair comparison. 

\begin{figure}[t]
    \centering
    \includegraphics[width=\linewidth]{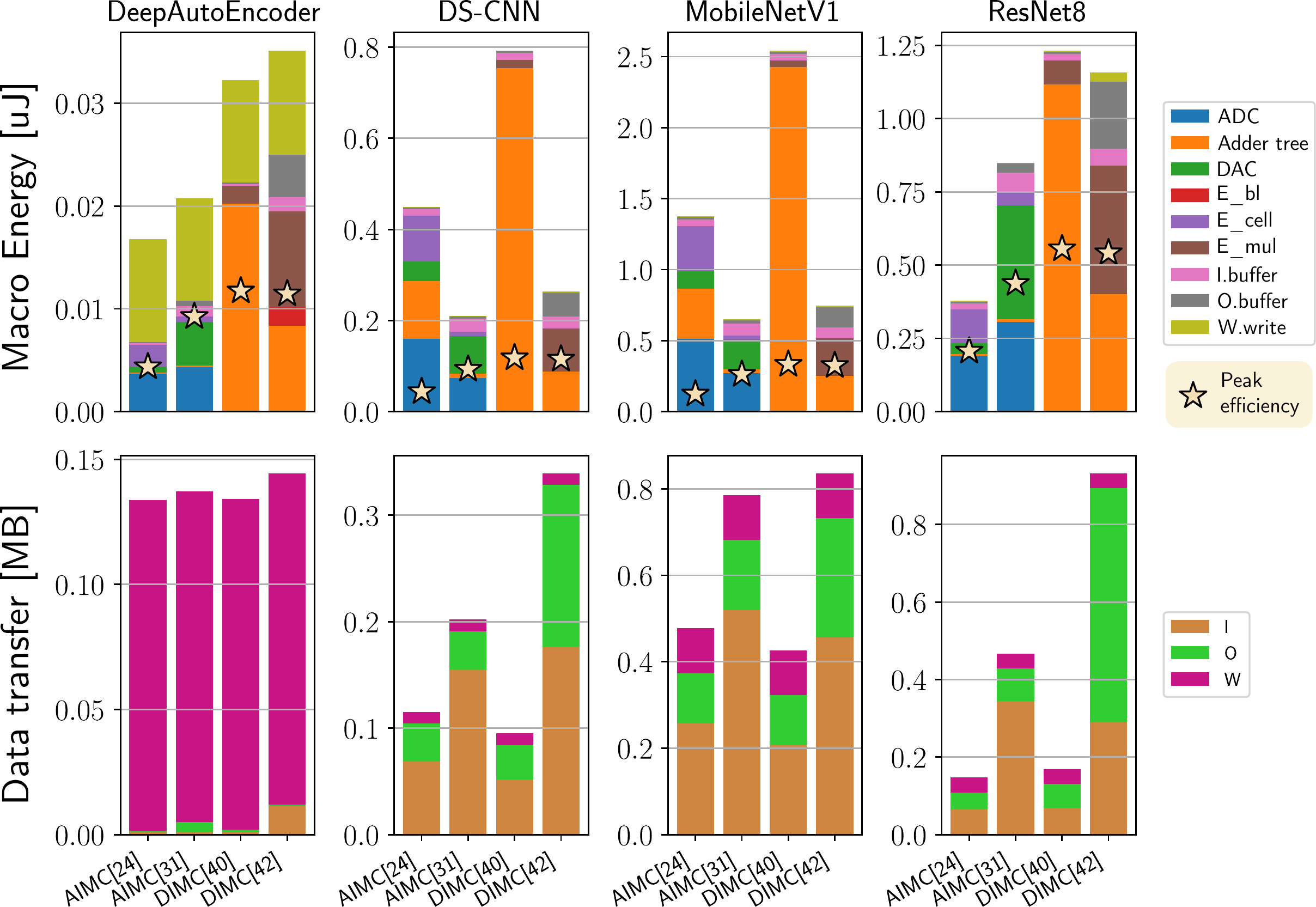}
    \label{fig:exp_1}
    \caption{Energy breakdown at macro level and required data traffic towards outer memory levels for selected IMC designs. Peak energy efficiencies highlight which designs more efficiently map the target workloads}
\end{figure}

The model with the parameters for the four specific designs has been integrated into the DSE framework ZigZag, allowing the tool to find the optimal spatial and temporal mapping for each architecture and each network layer. The result of the evaluations in terms of energy consumption at macro level and data transfer to/from the macro(s) are shown in Fig. \ref{fig:exp_1}. 

IMC architectures with large arrays are inherently best suited for networks which consist of layers that have high number of filter kernels and that require large accumulation over input channels and filter dimensions. These layers are usually convolutional ones, with filters larger than 1$\times$1, or fully connected ones. Those networks that exhibit such characteristics are the DeepAutoEncoder and the Resnet8. It can be seen in the energy breakdown in Fig. \ref{fig:exp_1} that AIMC architectures which feature large arrays are performing better than other designs; it has to be noted the benefits vanish if there is no reuse of the operands across computation cycles and weights have to be often rewritten. This occurs for the DeepAutoEncoder network: consisting of only fully connected layers (Fig. \ref{fig:fig_1} operator breakdown), no weight reuse can be obtained across computing cycles and weight data transfers increase energy consumptions. For ResNet8 nonetheless high efficiencies can be achieved, since temporally looping across feature map pixels amortizes the cost of writing the weights. On the other hand, smaller IMC arrays achieve high array utilizations but suffer from large overheads from the array peripherals.

On the opposite scenario, DS-CNN and MobilenetV1 networks are characterized by pointwise (i.e. convolutional layers with 1$\times1$ filters) and depthwise layers: the former require lower accumulation of partial results along the rows and the latter offer no reuse possibility for the activations across output channels mapped on the columns. This makes them unsuitable for IMC designs with large arrays as they cause heavy underutilization of the computational resources. For this, those architectures that feature smaller arrays but large number of macros perform better than their counterparts. Increasing the number of macros is a benefit as it allows spatial parallelization of different layer dimensions ($OX, OY, G$) beside the ones allowed in IMC (namely $C, K, FX, FY$ as shown in Sec. \ref{sec:02_background}). However, efficiency benefits at macro level come at a cost: input feature map pixels and partial accumulation values have to be fetched and stored from memory more often since less accumulation is carried out at macro level, as shown in the data transfer breakdown. 
Future works of design space exploration will focus on mitigating the feature map access overheads by placing extra levels of caching close to the computational macro in such cases, to further evaluate evaluate costs and area trade-offs.


\begin{table}[t]
\centering
\begin{tabular}{|l|c|c|c|c|c|c|}
\hline
\multicolumn{1}{|c|}{\textbf{Design}} & \textbf{R} & \textbf{C} & \textbf{Macros} & \textbf{\begin{tabular}[c]{@{}c@{}}Tech \\ {[}nm{]}\end{tabular}} & \textbf{V} & \textbf{\begin{tabular}[c]{@{}c@{}}A/W \\ bits\end{tabular}} \\ \hline
\textbf{AIMC \cite{aimc_2}}                 & 1152       & 256        & 1               & 28                                                                & 0.8        & 4b/4b                                                                                                                       \\ \hline
\textbf{AIMC \cite{aimc_7}}            & 64         & 32         & 8               & 28                                                                & 0.8        & 4b/4b                                                                                                                       \\ \hline
\textbf{DIMC \cite{dimc_1}}                & 256        & 256        & 4               & 22                                                                & 0.8        & 4b/4b                                                                                                                       \\ \hline
\textbf{DIMC \cite{dimc_3}}                 & 48         & 4          & 192             & 28                                                                & 0.8        & 4b/4b                                                                                                                       \\ \hline
\end{tabular}
\caption{Design characteristics of the compared architectures}
\label{tab:exp_1}
\end{table}
\section{Conclusions}

In this paper, we proposed a unified analytical cost model for AIMC and DIMC to 
be able to fairly compare across designs with different architectural parameters and technologies. 
We extract the parameters from a collection of published AIMC and DIMC designs across different technologies and realize a relative mismatch of about 15\%. Furthermore, integration of the developed model into the DSE framework ZigZag, allows to not only assess peak performances, but also actual workload performance when mapping real tinyMLperf benchmarks across diverse hardware IMC architectures. Future work will further deploy this model to assess the relative 
strengths and potential of AIMC and DIMC.

\section*{Acknowledgments}
This work has been supported by KU Leuven and received funding from the Flemish Government (Flanders AI Research Program) and the EU project CONVOLVE.
\bibliographystyle{IEEEtran}
\bibliography{refs}

\end{document}